\documentclass[12pt]{article}
\usepackage{amsmath,amsfonts,setspace,cancel,url,verbatim,soul,cite,bm}
\usepackage[hidelinks]{hyperref}
\usepackage{fullpage}
\usepackage{slashed}
\usepackage{cancel}
\usepackage{accents}
\newcommand{\be}{\begin{equation}}
\newcommand{\ee}{\end{equation}}

\newcommand\starred[1]{\accentset{\star}{#1}}
\numberwithin{equation}{section}
\newcommand\gh{{\rm gh}\;}
\def\tr{{\rm tr}}

\begin{document}
\begin{titlepage}
\begin{center}

%\hfill  DAMTP-2016-55 

\vskip 1.5cm

{\Large \bf A Weyl-invariant action for chiral strings and branes}

\vskip 2cm

 {\bf Alex S.~Arvanitakis\,${}^{1,2}$ }
 
 \vskip 15pt

{\em $^1$ \hskip -.1truecm
\em  Department of Applied Mathematics and Theoretical Physics,\\ Centre for Mathematical Sciences, University of Cambridge,\\
Wilberforce Road, Cambridge, CB3 0WA, U.K.\vskip 5pt }

\vskip .4truecm

{\em $^2$ \hskip -.1truecm
\em Department of Nuclear and Particle Physics,\\
Faculty of Physics, National and Kapodistrian University of Athens, \\
Athens 15784, Greece}
 \vskip 15pt

{email: {\tt A.S.Arvanitakis@damtp.cam.ac.uk}, {\tt alex.s.arvanitakis@gmail.com}} \\

 %and 
% %Chris D.~A.~Blair\,${}^2$} \\
% 
% 
% 
% \vskip 15pt
% 
% {\em $^1$ \hskip -.1truecm
% \em  Department of Applied Mathematics and Theoretical Physics,\\ Centre for Mathematical Sciences, University of Cambridge,\\
% Wilberforce Road, Cambridge, CB3 0WA, U.K.\vskip 5pt }
% 
% {email: {\tt A.S.Arvanitakis@damtp.cam.ac.uk}} \\
% 
% \vskip .4truecm

% {\em $^2$ \hskip -.1truecm
% \em Theoretische Natuurkunde, Vrije Universiteit Brussel, and the International Solvay Institutes, \\ 
%             Pleinlaan 2, B-1050 Brussels, Belgium \\ 
% \vskip 5pt }
% 
% {email: {\tt cblair@vub.ac.be}} \\
\end{center}

\begin{abstract}
% We introduce a sigma model lagrangian describing maps from a $p$-brane worldvolume into a symplectic space. The lagrangian is manifestly diffeomorphism and Weyl invariant despite the absence of a worldvolume metric. For $p=1$ this is a kind of chiral string lagrangian which can describe null (Schild) strings, (some?) ambitwistor strings and the recently introduced tensionless AdS twistor strings. We use this point of view to clarify issues of diffeomorphism invariance and metric dependence in these models; in particular we find that the stress energy tensor is classically BRST-exact in all cases (?). We also consider the implications for path integral quantisation and suggest a construction for the measure.
We introduce a sigma model lagrangian generalising a number of new and old models which can be thought of as chiral, including the Schild string, ambitwistor strings, and the recently introduced tensionless AdS twistor strings. This ``chiral sigma model'' describes maps from a $p$-brane worldvolume into a symplectic space and is manifestly diffeomorphism- and Weyl-invariant despite the absence of a worldvolume metric. Construction of the Batalin-Vilkovisky master action leads to a BRST operator under which the gauge-fixed action is BRST-exact; we discuss whether this implies that the chiral brane sigma model defines a topological field theory.

\end{abstract}
\end{titlepage}

\tableofcontents

\section{Introduction}

In this paper we will put forward a $p$-brane sigma model lagrangian and study its basic properties in a Batalin-Vilkovisky/BRST approach. For lack of a better name we will call this the ``chiral brane model'', even though properly speaking the model is ``chiral'' only for $p=1$. In that case, ``chiral'' means that the only derivative appearing in the action can be chosen to be a left-handed one, cf. the nomenclature of Siegel et al. in \cite{Siegel:2015axg,Huang:2016bdd}.

The $p=1$ version of the chiral brane model can be gauge-fixed to obtain a number of previously-known sigma models \cite{Schild:1976vq,Arvanitakis:2016vnp,Mason:2013sva,Geyer:2014fka,Claus:1999xr,Hohm:2013jaa}, most of which can be considered ``chiral'' and/or tensionless. We will discuss these models and their precise relation to the chiral brane model in section \ref{section_examples} later. Let us mention here that a common feature of the actions for these theories is a lack of manifest worldsheet diffeomorphism invariance, which formed the motivation for constructing the chiral sigma model. We will see shortly that the chiral sigma model enjoys manifest diffeomorphism invariance by virtue of a dynamical worldvolume vector field $v^a$, which is a new ingredient compared to the previously-known formulations of chiral strings, and whose introduction is accompanied by a new local scaling invariance. This ``Weyl invariance'' is closely analogous to that of the usual string except in that it acts on both worldvolume and target space 
at the same time.

The chiral brane action is:
\be
\label{def_S0}
S_0=S_0[z^A\,; v^a]=\frac{1}{2}\int d^{p+1}\sigma \: \left\{ \Omega_{AB} z^A \partial_a z^B v^a \right\}\,.
\ee
Here $\sigma^a, a=0,1,\dots p$ are $p$-brane worldvolume coordinates and the dynamical field $z^A=z^A(\sigma)$ describes the embedding of the worldvolume $\Sigma$ into a real symplectic vector space with symplectic form $\Omega_{AB}$ (assumed to be constant, antisymmetric and nondegenerate). The other dynamical field\footnote{``Dynamical'' in the sense that it is to be varied; one could call it auxiliary on account of the fact it appears as a lagrange multiplier, but we will refrain from doing so, as ``auxiliary'' might suggest it can be eliminated.},  $v^a=v^a(\sigma)$, carries a worldvolume vector index. We will take the worldvolume $\Sigma$ to be closed and accordingly drop all total derivatives inside integrals; one is thus inclined to think of the chiral brane theory as presented in this paper as a  Euclidean field theory, even though there is no worldsheet metric.

Later on we will generalise the action \eqref{def_S0} by adding the following interaction term involving a gauge field interacting with a bilinear in $z^A$ (where $M_{IAB}$ is constant):
\be
\label{def_gauged_interaction}
S_{\rm int}[z^A;A^I]=-\frac{1}{2} \int d^{p+1}\sigma \: \left\{ A^I M_{I A B} z^A z^B \right\}\,.
\ee
Those generalisations will be referred to as ``gauged'' chiral brane actions. The remarks we make in this section will also apply to the gauged models after appropriate modifications. One can also consider a complex version of \eqref{def_S0} which will be relevant for making contact with the ambitwistor string models in subsection \ref{section_ambitwistor}. In the complex, $p=1$ case, \eqref{def_S0} can be seen as an action describing a $\beta\gamma$ system. A final generalisation, which we will not treat in this paper, is to models with target space supersymmetry, for which we expect similar results to hold.

The action \eqref{def_S0} is manifestly invariant under worldvolume diffeomorphisms where $z^A$ transforms as a worldvolume scalar density of weight $+1/2$ and $v^a$ transforms as a vector field (of weight zero). Perhaps surprisingly, it is also invariant under a \emph{local} Weyl transformation if $z^A$ transforms with weight $-1$ and $v^a$ transforms with weight $+2$. The infinitesimal transformations (with parameters $\xi^a=\xi^a(\sigma)$ and $\omega=\omega(\sigma)$ respectively) read
\begin{align}
\delta_\xi z^A &= \xi^a \partial_a z^A + \frac{1}{2}(\partial_a \xi^a) z^A\,,& \qquad \delta_\xi v^a &= \xi^b\partial_b v^a- v^b\partial_b \xi^a \\
\delta_\omega z^A &= -\omega z^A\,,& \qquad \delta_\omega v^a &= + 2 \omega z^A\,.
\end{align}
Invariance under local Weyl transformations follows from the antisymmetry of $\Omega_{AB}$.

It is not entirely clear whether the model can be consistently defined on arbitrary symplectic manifold target spaces: while one can always cover any symplectic manifold by Darboux charts where the symplectic form is constant, whether such local descriptions can be sewn together in a consistent way is not obvious. We will anyway not consider this generalisation in this paper.

A peculiarity of the gauge transformations just presented is that one can combine them to shift the density weight of the dynamical fields under worldvolume diffeomorphisms. A Weyl transformation with parameter
\be
\label{density_weight_redefinition}
\omega= -x \partial_a \xi^a\,;\qquad x\in \mathbb{R}
\ee
can be combined with an infinitesimal diffeomorphism to shift the weight of $z^A$ by $x$ and that of $v^a$ by $-2x$. Thus the form of the transformations we exhibited above is not unique. A second peculiarity is that an infinitesimal diffeomorphism with parameter $\xi^a=v^a$ always vanishes on-shell. We might thus be concerned that the gauge transformations are reducible (i.e. that there are ``gauge transformations of gauge transformations''). One could, however, guess that this cannot be the case on account of the fact the space of on-shell vanishing gauge transformations is one dimensional and is spanned by $v^a$.

To clarify these issues we will analyse the model in the Batalin-Vilkovisky (BV) formalism \cite{Batalin:1984jr} in section \ref{section_BRST}, with the result that the gauge transformations close off-shell and are irreducible. The shift in $x$ described in the previous paragraph will be realised by a canonical transformation in the BV sense, and thus theories for different values of $x$ are equivalent; as such we will generally consider the theory with $x=0$ only.

A fact that falls out of the BV analysis is that $S_0$ is trivial in the gauge-invariant BRST cohomology (in the terminology of \cite{Henneaux:1995ex}). After gauge-fixing in the usual way a little more work leads to a gauge-fixed action which is a total BRST variation, i.e. trivial in the BRST cohomology. One would then conclude that the chiral brane action describes a cohomological field theory\footnote{Such theories are more often called ``topological'' but we prefer to use this more precise language --- after all the chiral brane action is already ``topological'' in that does not depend on a worldvolume metric.} in the sense of \cite{Witten:1990bs}. However we will not be able to prove whether the action remains trivial in the BRST cohomology after certain ghost variables have been eliminated. Furthermore, the theory described by $S_0$ clearly possesses local degrees of freedom and it would thus be highly counterintuitive if it were cohomological at the same time. This point is further discussed in subsection \ref{subsection_unbelievers}. For these 
reasons the status of the chiral brane theory as a cohomological field theory 
is somewhat unclear.

\section{Batalin-Vilkovisky master action and BRST transformations}
\label{section_BRST}

The Batalin-Vilkovisky (BV) formalism \cite{Batalin:1984jr} is an approach for handling gauge theories where the algebra of gauge transformations only closes on-shell and/or the gauge transformations are reducible, and more specifically it is often used to construct BRST operators and BRST-invariant actions for such theories. For the chiral brane sigma model we will also see how it clarifies the structure of the gauge transformations. We will give a very brief outline of the formalism in the next paragraph and refer to the review \cite{Henneaux:1989jq} for details. We also profited from the discussion in \cite{Craps:2005wk}.

The central object in the BV approach is the ``master action''. This is constructed from the original action $S_0$ (\eqref{def_S0} in this case) in a number of steps. One first takes the original gauge parameters ($\xi^a,\omega$) and introduces a set of corresponding ghost fields ($c_\xi^a, c_\omega$), which are defined to have ghost number $\gh c_\xi^a=\gh c_\omega=1$. The original fields ($z^A,v^a$) are defined to have ghost number zero. It is customary to refer to the collection of original fields plus ghost fields as just ``fields''. For each field $\phi$ one then introduces a corresponding \emph{antifield} $\starred \phi$ of ghost number $\gh \starred \phi=-\gh \phi -1$ which is conjugate to $\phi$ in the sense
\be
\label{def_antibracket}
(\phi(\sigma),\starred \phi(\sigma'))=\delta^{p+1}(\sigma-\sigma')\,.
\ee
The (anti)fields are bosonic or fermionic depending on their ghost number modulo 2 (assuming we started with a purely bosonic theory). The bracket $(\--,\--)$ is known as the antibracket and is graded anticommutative in its two arguments with respect to the grading by ghost number. It also satisfies a modified version of the Leibniz rule which can be summarised by saying the bracket ``carries ghost number +1''. The master action $S_{\rm BV}=S_{\rm BV}[\phi,\starred \phi]$ is then defined to be a ghost number zero functional which solves the master equation
\be
(S_{\rm BV},S_{\rm BV})=0
\ee
such that $S_{\rm BV}[\phi,\starred \phi=0]=S_0$. If the gauge transformations are reducible one needs more ghosts (with corresponding antifields) than described above but we will see that the general case is not relevant for this paper.

The master action encodes the gauge transformations in its antifield dependence. Specifically it is easy to see that the terms linear in antifields of the original fields (i.e. linear in $\starred z_A, \starred v_a$) must be proportional to the original infinitesimal gauge transformations with gauge parameter replaced by the corresponding ghost. Higher-order terms in antifields are only present if the gauge algebra only closes on-shell, and more ghost fields than described above are needed if the gauge transformations are reducible.

For the ungauged chiral brane sigma model action \eqref{def_S0} a master action solving the master equation is
\be
S_{\rm BV}=S_0+S_1
\ee
with
\be
S_0=S_0[z^A\,; v^a]=\frac{1}{2}\int \: \left\{ \Omega_{AB} z^A \partial_a z^B v^a \right\}
\ee
and
\begin{align}
S_1=\int \: \bigg\{ \starred z_A \left( c_\xi^a \partial_a z^A +\frac{1}{2} \partial_a c^a_\xi z^A - c_\omega z^A \right) + \starred v_a (c^b_\xi \partial_b v^a -v^b\partial_b c^a_\xi +2 c_\omega v^a ) \nonumber\\
-\starred c^\xi_a c^b_\xi \partial_b c^a_\xi - \starred c_\omega c^b_\xi\partial_b c_\omega\bigg\}\,,
\end{align}
where here and henceforth $\int=\int d^{p+1}\sigma$ unless noted. Notice that both $S_0$ and $S_1$ are real because we are using the atypical complex conjugation convention (also used in \cite{Arvanitakis:2016vnp,Arvanitakis:2016wdn}) that sends $\psi_1\psi_2\to \bar \psi_1 \bar \psi_2$ if $\psi_1,\psi_2$ are fermionic.
From the fact the master action is linear in antifields we see that the gauge transformations close off-shell and that they are irreducible. 
The coefficients of the two terms on the last line are fixed by the master equation.

The master equation is equivalent to the invariance of $S_{\rm BV}$ under infinitesimal BRST transformations defined as
\be
\delta_{\rm BRST} \Phi\equiv (\Phi , S_{\rm BV} \Lambda)\,.
\ee
Here $\Lambda$ is a constant anticommuting parameter of ghost number $\gh \Lambda=-1$, where the ghost number of $\Lambda$ was chosen so that $\delta_{\rm BRST}$ satisfies the Leibniz rule. We have
\be
\delta_{\rm BRST}^2\Phi\equiv\delta_{{\rm BRST};\Lambda_1} (\delta_{{\rm BRST};\Lambda_2}\Phi)=0
\ee
as a consequence of the master equation and a super Jacobi identity satisfied by the antibracket. These BRST transformations are related to but not the same as the ones usually employed in the context of BRST quantisation after the antifields are eliminated.

One then defines observables to be functionals $F$ which are BRST-closed $(\delta_{\rm BRST} F=0)$ modulo BRST-exact ones, i.e. we identify
\be
F\sim F' + \delta_{\rm BRST} G\,.
\ee
The cohomology thus defined is called the \emph{gauge-invariant BRST cohomology}, to contrast it with the usual BRST cohomology (also known as gauge-fixed BRST cohomology) which we will focus on later.

Usually the classical action $S_0$ is an observable (it is BRST-closed by virtue of its gauge-invariance). The chiral brane theory however happens to enjoy the peculiar property that $S_0$ is BRST-exact! This can in fact be verified almost by inspection, as the BRST variation of $\starred v_a v^a$ will include a term proportional to the original lagrangian. In fact after a short calculation we find
\be
(-\starred v_a v^a , S_{\rm BV})= \frac{1}{2} \Omega_{AB} z^A \partial_a z^B v^a + \partial_b(c^b_\xi \starred v_a v^a)\,.
\ee
Thus after integrating over the worldvolume (assumed closed) we find that $S_0$ is BRST-trivial after dropping the boundary term. More precisely we have just found that $S_0$ is BRST-trivial in the gauge-invariant BRST cohomology of local functionals.  The relation between this cohomology and the gauge-fixed BRST cohomology we are actually interested in can be subtle (see e.g. \cite{Henneaux:1995ex}), so we will revisit this issue after constructing the BRST operator for the gauge-fixed cohomology.

The solution to the master equation is not unique. For the chiral brane sigma model this ambiguity includes the ambiguity in the density weights noted in the Introduction. The BV action above was written down for a specific choice of weights, so it stands to reason there should exist equivalent BV master actions corresponding to each consistent weight choice. We point out that they can be obtained using the canonical transformation defined by the fermion
\be
\label{def_canonical_transformation_density_weight}
\Psi_x=x\int \starred c_\omega \partial_a c^a_\xi\,, \qquad x\in \mathbb{R}
\ee
where canonical transformations act in the standard way:
\be
e^{\Psi_x}\Phi= \Phi + (\Psi_x,\Phi) + \dots\,,\qquad \text{($\Phi$ is any (anti)field).}
\ee
$\Psi_x$ is referred to as a fermion as it must have ghost number -1 if the transformed action is to have ghost number zero.
This transformation produces the shifts
\be
c_\omega\to c_\omega- x \partial_a c^a_\xi\,,\qquad \starred c^\xi_a \to \starred c^\xi_a - x \partial_a \starred c_\omega
\ee
in the BV action. Because this $\Psi_x$ transformation is canonical, the BV action thus modified satisfies the master equation for any $x$.

Since these shifts in $x$ are canonical transformations we know that the quantum theory will be independent of the choice of $x$ (in the absence of BRST anomalies; see e.g. \cite{Henneaux:1989jq} section 8.9). Besides $x=0$, another apparently equally natural choice is $x=-1/2$, for which the fields $z^A$ transform as worldvolume scalars under diffeomorphisms, while $v^a$ turns into a vector density of weight 1. However, it seems that $x=0$ is singled out if one demands equivalence with the canonical Hamiltonian formalism (after a fairly innocent-looking choice of partial gauge fixing), which will be discussed in section \ref{section_examples}.

\subsection{Gauge fixing and BRST exact action (ungauged model)}

The BV action $S_{\rm BV}$ always has a number of local fermionic invariances and thus requires gauge fixing. These can always fixed by setting the antifields to zero, however doing so in the original BV action simply leads us back to the original action, which has troublesome gauge invariances of its own. The way out is to first perform a canonical transformation so that what remains after the antifields vanish has no gauge invariances. In practice, one does not always eliminate all gauge invariances this way (consider e.g. the usual string in conformal gauge for low-genus worldsheets) but the end action tends to be rather more amenable to path integral methods by virtue of possessing BRST invariance. We will simply view the BV apparatus as a way to obtain the BRST transformations.

We gauge fix in the standard way by introducing the ``non-minimal sector'' fields $\pi_a$ and $b_a$ with antifields $\starred b^a$ and $\starred \pi^a$ respectively, with ghost numbers $\gh \pi_a=0$ and $\gh b_a=-1$. The field $b_a$ is confusingly known in the literature as the ``antighost'', which however should not be confused with the $c^a_\xi$ ghost antifield $\starred c_a^\xi$. The solution of the master equation is then modified by the addition of the term $\starred b^a \pi_a$. Thus the starting point of the gauge fixing procedure is the ``non-minimal action''
\be
S_{\rm BV} + \int \starred b^a \pi_a\,,
\ee
which clearly satisfies the master equation whenever $S_{\rm BV}$ does. We then consider the gauge fixing fermion 
\be
\label{gaugefixing_fermion}
\Psi=\int b_a (v^a-\tilde v^a)
\ee
where $\tilde v^a\neq 0$ is nondynamical and its components are assumed to be constant (this gauge can always be reached locally on the worldvolume as long as $v^a$ does not vanish). Without loss of generality we can thus use coordinates where $\tilde v^a=(1,0,\dots 0)^T$. The canonical transformation generated by $\Psi$ amounts to the shifts
\be
\starred b^a \to \starred b^a + (v^a-\tilde v^a)\,,\qquad \starred v_a \to \starred v_a + b_a
\ee
in the non-minimal action. We thus obtain
\be
S_{{\rm BV};\Psi}=S_0  + \int b_a (c^b_\xi \partial_b v^a -v^b\partial_b c^a_\xi +2 c_\omega v^a )+ \int (v^a-\tilde v^a)\pi_a +\int \starred b^a \pi_a +S_1
\ee
which is related to the gauge-fixed action $S_{\rm gf}$ by setting all antifields (i.e. all starred fields) to zero:
\be
S_{\rm gf}[\phi]=S_{{\rm BV};\Psi}[\phi,\starred \phi=0]\,.
\ee
This has the effect of making the last two terms of $S_{{\rm BV};\Psi}$ vanish.

The upshot of this so far standard analysis is that the gauge-fixed action inherits the BRST invariance of $S_{{\rm BV};\Psi}$ by construction: This is obvious if we rewrite
\be
\label{Sgaugefixed_brst1structure}
S_{\rm gf}[\phi]=S_0 + (\Psi, S_{{\rm BV};\Psi})|_{\starred \phi=0}\,.
\ee
For an irreducible gauge theory such as the chiral brane it is well-known that the following BRST transformations 
\be
\delta_{\rm BRST1}\phi \equiv (\phi,S_{{\rm BV};\Psi}\Lambda)|_{\starred \phi=0}
\ee square to zero off-shell and thus the above expresion is invariant. We have named these transformations ``BRST1'' for reasons to become apparent.

We will now modify these BRST transformations. The modification is by the following fermionic ``trivial transformation'' (in the sense that it vanishes on-shell) with constant anticommuting parameter $\Lambda$ of ghost number $-1$:
\be
\label{def_trivial}
\delta_{\rm{trivial}} v^a= \frac{\delta S_{{\rm BV};\Psi}}{\delta b_a} \Lambda \,, \qquad \delta_{\rm{trivial}} b_a =\frac{\delta S_{{\rm BV};\Psi}}{\delta v^a} \Lambda\,,
\ee
where the functional derivative is defined as $\delta S=\int \delta \phi \;(\delta S/\delta \phi)$. This is an invariance of $S_{{\rm BV};\Psi}$. The modified BRST transformations acting on $S_{\rm gf}$ are then defined as
\be
\label{def_BRST_modified}
\delta_{\rm BRST2}\phi\equiv((\phi, S_{{\rm BV};\Psi}\Lambda)-\delta_{\rm trivial} \phi)|_{\starred \phi=0}\,,\qquad \text{($\phi$ is any field).}
\ee
This new BRST variation (which we named ``BRST2'' to avoid confusion with the one defined in the previous paragraph) automatically satisfies $\delta_{\rm BRST2}^2=0$ on-shell as a consequence of the master equation and the fact $\delta_{\rm trivial}$ vanishes on-shell. However for this theory a direct calculation shows that in fact we have $\delta_{\rm BRST2}^2=0$ \emph{off-shell} as well, which will be imporant in what follows.

We list the BRST transformations in full in the appendix. To verify BRST invariance, we only need calculate
\be
\delta_{\rm BRST2} v^a=0\,, \qquad \delta_{BRST2} \pi_a=0
\ee
and
\be
\delta_{\rm BRST2} b_a= \left(-\frac{1}{2}\Omega_{AB} z^A \partial_a z^B + \partial_b(b_a c^b_\xi)+ b_b \partial_a c^b_\xi -2 b_a c_\omega\right)\Lambda.
\ee
The fact $v^a$ is BRST-closed partly motivated our choice of BRST variation). Another motivation comes from the fact that with this choice of BRST operator we have
\be
\label{Sgaugefixed_brst2structure}
S_{\rm gf}\Lambda= \delta_{\rm BRST2} \left(\int -b_a v^a\right) + \int (v^a-\tilde v^a)\pi_a \Lambda\,.
\ee
Both terms on the right-hand side are individually BRST-invariant.

The fields $\pi_a$ and $v^a$ can be integrated out together. This imposes the gauge condition $v^a=\tilde v^a={\rm const.}$ in $S_{\rm gf}$, leaving
\be
S_{{\rm gf}'}=\int \left\{\frac{1}{2} \Omega_{AB}z^A \tilde v^a\partial_a z^B  - b_b \tilde v^a \partial_a c^b_\xi +2 b_a \tilde v^a c_\omega\right\}
\ee
after dropping a boundary term. The BRST variation after $\pi_a$ and $v^a$ have been eliminated is still nilpotent off-shell, as was verified by direct calculation. As the BRST variations are the same, we find:
\be
S_{{\rm gf}'} \Lambda = \delta_{\rm BRST2} \left(\int -b_a \tilde v^a\right)\,.
\ee
We have thus recovered the result that the action is BRST-exact, now in the gauge-fixed BRST cohomology defined in terms of the modified BRST variation \eqref{def_BRST_modified} (BRST2) above.

The BRST variation BRST2 after $\pi_a$ and $v^a$ have been eliminated is still nilpotent off-shell. For this reason, whenever we mention the ``gauge-fixed'' action and BRST variations in the rest of the paper (e.g. in the context of the gauged chiral brane models we will discuss shortly) we will be referring to $S_{{\rm gf}'}$ (i.e. the gauge-fixed action with $\pi_a$ and $v^a$ eliminated) and the variation BRST2 \eqref{def_BRST_modified}, except as noted.

\subsubsection{A ``topological'' theory with local degrees of freedom?}
\label{subsection_unbelievers}
To summarise: the claim so far is that starting from the action \eqref{def_S0} and gauge fixing the dynamical variable $v^a$ to a nonvanishing constant $\tilde v^a$ in a BRST procedure leads to the gauge-fixed, BRST-invariant action
\be
S_{{\rm gf}'}[z^A,b_a,c^a_\xi,c_\omega]=\int \tilde v^a \left\{\frac{1}{2} \Omega_{AB}z^A \partial_a z^B  - b_b  \partial_a c^b_\xi +2 b_a  c_\omega\right\}
\ee
which is BRST-exact in the sense
\be
S_{{\rm gf}'} \Lambda = \delta_{\rm BRST} \left(\int -b_a \tilde v^a\right) \qquad \text{($\Lambda$ is the constant BRST transformation parameter)}
\ee
and where the BRST transformations $\delta_{\rm BRST}\equiv\delta_\Lambda$ defined in \eqref{def_BRST_modified} (the ones named ``BRST2'' in the previous section) satisfy
\be
\delta_{\Lambda_2}(\delta_{\Lambda_1} z^A)=\delta_{\Lambda_2}(\delta_{\Lambda_1} b_a)=\delta_{\Lambda_2}(\delta_{\Lambda_1} c_\xi^a)=\delta_{\Lambda_2}(\delta_{\Lambda_1} c_\omega)=0
\ee
identically, i.e. they square to zero \emph{off-shell}. The explicit formulas for the BRST variations are \eqref{BRST_z}, \eqref{BRST_ca}, \eqref{BRST_comega}, and \eqref{BRST_ba}
(\emph{excluding} the terms with $IJK$ indices for now since we have not considered the generalisation to the gauged model yet). Their off-shell nilpotence was verified by hand and also using the computer algebra programme Cadabra (v. 1.39) \cite{DBLP:journals/corr/abs-cs-0608005,Peeters:2007wn}.

A perhaps disturbing feature\footnote{I am grateful to Paul Townsend for pointing this out to me.} of that calculation is that while the full gauge-fixed action was found to be BRST-exact, it is not clear whether $S_0$ by itself is. One might think that this might necessarily be the case: after all, in a standard approach the BRST-invariant gauge-fixed action ($S_{\rm gf}$ or $S_{\rm gf'}$ in the previous section) is obtained from the original action $S_0$ by the addition of a BRST-exact term. In fact, this is exactly what we found, for the BRST variation ``BRST1'' of the previous section, in formula \eqref{Sgaugefixed_brst1structure} which we reproduce here in more compact notation:
\be
S_{\rm gf}\Lambda=S_0 \Lambda + \delta_{\rm BRST1}\Psi\,.
\ee
where $\Psi$ is the gauge-fixing fermion \eqref{gaugefixing_fermion}, $\Lambda$ is the constant BRST transformation parameter, and we are using the gauge-fixed action $S_{\rm gf}$ where the fields $\pi_a$ and $v^a$ have not been eliminated yet. The loophole lies in that when we modified the BRST variations from BRST1 to BRST2 by terms proportional to equations of motion, the formula analogous to the above was modified to \eqref{Sgaugefixed_brst2structure}, reproduced here:
\be
S_{\rm gf}\Lambda= \delta_{\rm BRST2} \left(\int -b_a v^a\right) + \int (v^a-\tilde v^a)\pi_a \Lambda\,.
\ee
We see that under BRST1, $S_0$ is BRST-closed while the term $(v^a-\tilde v^a)\pi_a$ plus fermions is BRST-exact, while under BRST2 $S_0$ plus fermions is BRST-exact while $(v^a-\tilde v^a)\pi_a$ by itself is only BRST-closed. The existence of both off-shell nilpotent BRST variations BRST1 and BRST2 is of course a special feature of the chiral brane theory and appears to be intimately related to the worldvolume vector field $v^a$:
\be
\delta_{\rm BRST1} b_a= \pi_a \Lambda \approx\left(-\frac{1}{2} z^A \Omega_{AB} \partial_a z^B +\dots\right)\Lambda = \delta_{\rm BRST2} b_a\,.
\ee
The first equality represents the usual form of the BRST variation $\delta_{\rm BRST1}b_a$ of an antighost $b_a$. Using the $v^a$ equation of motion, this is equal to $\delta_{\rm BRST2}b_a$.

The claim that the action is BRST-exact might come across as counterintuitive, for the following reason: consider the original action \eqref{def_S0} and partially gauge-fix $v^a=(v^0,v^i)=(1,-\rho^i)$ (in coordinates $\sigma^a=(t,\sigma^i)$, $i=1,\dots p$) to get
\be
\label{def_hamiltoniangaugefixing_ungaugedaction}
\int dt d^p\sigma \: \left\{ \frac{1}{2}\Omega_{AB} z^A \dot z^B  - \rho^i \left(\frac{1}{2} \Omega_{AB} z^A \partial_i z^B \right) \right\}\,.
\ee
If the real symplectic vector space with coordinates $z^A$ is of dimension $2d$, then after an arbitrary choice of $d$ positions and $d$ momenta we see that the action \eqref{def_hamiltoniangaugefixing_ungaugedaction} takes the standard form for a constrained Hamiltonian system with $p$ spatial diffeomorphism constraints enforced by the lagrange multipliers $\rho^i$ and therefore describes $d-p$ configuration space degrees of freedom (since the constraints displayed are first-class).

The conclusion is that the chiral brane theory with the BRST operator BRST2, is apparently a cohomological theory with local degrees of freedom! A priori this sounds like a contradiction in terms, especially since the action in fact satisfies the stronger property of being a BRST variation and one then expects to be able to set it to zero by deforming the gauge condition. Indeed, setting $\tilde v^a=0$ makes the action vanish. However this is not a sensible gauge condition: recall that $v^a$ transforms as a worldvolume vector field under diffeomorphisms and is rescaled by Weyl transformations, so it cannot be set to zero in small neighbourhoods of any point where it is nonvanishing, and moreover the condition $v^a=0$ does not actually fix the gauge. The best one can do is set $v^a=\tilde v^a=(1,0,0,\dots)^T$.

Another relevant observation is that we have not been able to prove that the gauge-fixed action remains BRST-trivial after eliminating certain ghost fields by their own equations of motion: if we set $v^a=\tilde v^a=(1,0,0,\dots)^T$ in coordinates $\sigma^a=(t,\sigma^i), t=\sigma^0$ as before, we can write the gauge-fixed action as
\be
S_{{\rm gf}'}=\int dt d^p\sigma \:\left\{ \frac{1}{2} \Omega_{AB}z^A \dot z^B - b_0 \dot c^0_\xi - b_i \dot c^i_\xi + 2 b_0 c_\omega \right\}
\ee
where we have split $b_a=(b_0,b_i)$ and $c_\xi^a=(c^0_\xi,c^i_\xi)^T$. Then $b_0$ and $c_\omega$ can be jointly eliminated to set $b_0=0$ and $c_\omega = \dot c^0_\xi/2$ and leave
\be
S_{{\rm gf}''}=\int dt d^p\sigma \:\left\{ \frac{1}{2} \Omega_{AB}z^A \dot z^B  - b_i \dot c^i_\xi \right\}. 
\ee
Since $b_0=b_a\tilde v^a$ has been set to zero there appears to be no candidate expression which we could vary to obtain $S_{{\rm gf}''}$.

The status of the chiral brane theory as a cohomological field theory thus appears to depend on the choice of ghost fields to be integrated over in the path integral. This is ultimately a choice of path integral measure. It is, in particular, possible that consistency requires a path integral measure which involves integrations over extra fermion variables. This was actually found to be the case in \cite{Dedushenko:2010zn} for the $p=0$ case of the chiral brane, which is --- in the ungauged case --- simply a particle on phase space with vanishing Hamiltonian. It is therefore conceivable that the full set of ghosts ($c_\omega, b_a, c^a_\xi$) are required by consistency, as opposed to the minimal set $(b_i,c^i)$; in that case whether $c_\omega$ and $b_a\tilde v^a$ can be integrated out should be investigated carefully. Such quantum considerations would, however, have to be the subject of another paper.

Before ending this subsection we note that similar non-conclusions follow for the gauged chiral brane model, which we will consider next.

\subsection{Gauge fixing and BRST exact action (gauged model)}

The gauged chiral brane model has bosonic action
\be
\label{def_gauged_bosonic_action}
S'_0[z^A;v^a,A^I]=S_0+S_{\rm int}=\frac{1}{2}\int d^{p+1}\sigma \: \left\{ \Omega_{AB} z^A \partial_a z^B v^a  - A^I M_{I A B} z^A z^B\right\}\,.
\ee
We will assume the matrices $M_{I A B}$ are constant and symmetric in $AB$. They are to be thought of as determining a Lie algebra of gauge transformations acting on $z^A$, for which $A^I$ is the gauge field and lagrange multiplier. The most straightforward way to see this is to make an arbitrary definition of worldvolume time $t$ and gauge fix $v^a\partial_a=\partial_t$. When this is the case we have an ordinary phase space action in Hamiltonian form for the phase space spanned by $z^A$ with canonical Poisson brackets $\{z^A,z^B\}=\Omega^{AB}$. Then
\be
T_I\equiv M_{I A B} z^A z^B
\ee
define a number of constraints on the phase space spanned by $z^A$, and those constraints are first-class in the sense of Dirac (and thus correspond to gauge transformations) if the Poisson bracket algebra closes, i.e. (the factor of 2 is conventional)
\be
\label{def_f}
\{ T_I, T_J\}=2 f_{IJ}^{}{}^K T_K \iff \frac{1}{2} M_{KAB} f_{IJ}^{}{}^K= M_{I A C}\Omega^{CD} M_{J BD}\,.
\ee
The quantities $f_{IJ}^{}{}^K$ are thus constant when the algebra closes and are interpreted as the structure constants of the Lie algebra of the $T_I$ constraints. The Jacobi identity for the $f_{IJ}^{}{}^K$ is implied by that of the Poisson bracket.

At this point one could construct the BV master action for this gauged model and proceed as before to obtain the gauge-fixed BRST-invariant action, analogous to $S_{{\rm gf}'}$ above. In that derivation however, the fact $\delta_{\rm BRST}^2=0$ off-shell was only shown by  direct calculation. We therefore found it more economical to simply deform the BRST transformations BRST2 \eqref{def_BRST_modified} of the ungauged model by adding terms involving the new fields and ghosts (corresponding to the new gauge invariances in the gauged theory), and then constrain the relative coefficients by demanding that the new transformation be nilpotent, rather than go through the convoluted procedure of the previous subsection.

Let us sketch how this works. In the gauge $A^I=0$, which is an admissible gauge choice for the above action since the gauge transformation of $A^I$, in contrast to that of $v^a$, is inhomogeneous, the number of possible deformation terms is rather limited: the new terms can only depend on $c^I$ (the ghost for the gauge transformations generated by $T_I$), the corresponding antighost $b_I$ (of ghost number $-1$), and the constant matrices $M_{IAB}$, $f_{IJ}^{}{}^K$, and $\Omega_{AB}$ as well as its inverse $\Omega^{AB}$. At the same time we constrain the BRST variations to be quadratic in the fields because we expect the putative BV action to be purely cubic, like in the ungauged theory. This fact along with some ghost number counting implies that the candidate deformation terms in the BRST variation of the original antighost, $b_a$, are $(b_I \partial_a c^I)\Lambda$ and $(\partial_a b_I c^I)\Lambda$. Assuming that such an off-shell nilpotent BRST variation exists we can therefore immediately write down a 
gauge-fixed BRST-exact action:
\begin{align}
\label{def_gauged_BRST_action}
S=\int \left\{\frac{1}{2} \Omega_{AB}z^A \tilde v^a\partial_a z^B  - b_b \tilde v^a \partial_a c^b_\xi +2 b_a \tilde v^a c_\omega - b_I \tilde v^a\partial_a c^I\right\}\,,\\ \delta_{BRST}\left(-\int b_a\tilde v^a\right)= S \Lambda\,.
\end{align}
The coefficient on the last term of \eqref{def_gauged_BRST_action} is fixed by the detailed calculation of $\delta_{\rm BRST}$ outlined in the appendix.

The only conditions which must hold in order that $\delta_{\rm BRST}$ squares to zero off-shell were found to be
\begin{align}
\label{BRST_conditions}
\Omega^{AC}\Omega_{BC}=\delta^A_B\,,\qquad \frac{1}{2} M_{KAB} f_{IJ}^{}{}^K= M_{I A C}\Omega^{CD} M_{J BD}\,, \qquad
f_{[IJ}^{}{}^L f_{K]L}^{}{}^M=0\,.
\end{align}
The first two are the definitions of $\Omega^{AB}$ and $f_{IJ}^{}{}^K$ respectively, while the last is the Jacobi identity for $f_{IJ}^{}{}^K$. As the $f_{IJ}^{}{}^K$ are the same structure constants (as in \eqref{def_f}) appearing in the canonical Hamiltonian analysis we outlined above , we find that \emph{any gauged chiral brane sigma model has a BRST-exact action} (and thus describes a cohomological field theory, subject to the caveats of subsection \eqref{subsection_unbelievers}) \emph{whenever the constraints $T_I$ close into a first-class constraint algebra}. One can usually determine this closure at a glance. It is then easy to find examples of gauged chiral $p$-brane sigma models in the literature, at least for the case $p=1$.

\section{$p=1$: Chiral string examples}
\label{section_examples}
Here we will show out how the bosonic gauged chiral string action \eqref{def_gauged_bosonic_action} is related to a number of different first order string actions which have previously appeared in the literature in various contexts. In all cases this will be done by an appropriate choice of the target symplectic space alongside a choice of gauge-fixing for the components of $v^a$, and possibly redefinitions of $z^A$.

\subsection{Tensionless Minkowski (Schild) string}
\label{section_schild}
A lagrangian for the tensionless string in Minkowski space was first proposed by Schild approximately 40 years ago \cite{Schild:1976vq}. We will look at a version given later by Lindstrom, Sundborg and Theodoridis \cite{Lindstrom:1990qb} in its phase space form:
\be
\label{def_tensionless_Mink}
S[X^\mu, P_\mu; \lambda,\rho]=\int dt d\sigma \: \left\{ \partial_t X^\mu P_\mu - \lambda \left( P^2\right) -  \rho (\partial_\sigma X^\mu P_\mu)\right\}\,,
\ee
where $\lambda$ and $\rho$ are lagrange multipliers. This is related to the action for the usual (tensile) string through the replacement $\lambda \left( P^2\right)\to \lambda\left( P^2 +(T\partial_\sigma X)^2\right)$, where $T$ is the tension.

This tensionless string action is obtained from \eqref{def_gauged_bosonic_action} if we set
\begin{align}
z^A=\begin{pmatrix} X^\mu \\ \phi^{-1}P_\mu\end{pmatrix}\,,\quad \Omega_{AB}= \begin{pmatrix} 0 & -1 \\ 1 & 0 \end{pmatrix}\,,\quad v^a=\begin{pmatrix} v^t\\v^\sigma\end{pmatrix} = \phi\begin{pmatrix} 1\\-\rho\end{pmatrix}\,,\quad
A^I=2\lambda\phi^2\,,\nonumber\\  M_{IAB}z^Az^B= \phi^{-2}P^2\,,
\end{align}
and integrate by parts, where $\phi$ (which will drop out) is constrained by the requirement
\be
\partial_a v^a=0\,.
\ee
We can view this requirement as a gauge-fixing condition for the local Weyl transformations if $v^a$ is assumed to transform as a worldsheet vector density of weight $+1$  so that $\partial_a v^a=0$ is a diffeomorphism invariant condition. With that assumption $z^A$ and thus $X^\mu$ must transform as a scalar; then the remaining gauge transformations match fully if we assume $P_\mu$ transforms as a scalar density of weight $+1$.

We note that $v^a$ can be identified up to a Weyl transformation with the vector density $V^a$ (of weight $+1/2$) appearing in the tensionless string action after the momenta $P_\mu$ have been eliminated:
\be
\label{def_LST_Action}
S[X^\mu;V^a]=\frac{1}{2}\int d^2\sigma \: \left\{V^a V^b\partial_a X^\mu \partial_b X_\mu \right\}\,,\qquad V^a =\frac{1}{2\sqrt{\lambda}} \begin{pmatrix} 1\\-\rho\end{pmatrix}\,.
\ee
In \cite{Lindstrom:1990qb}, after comparison with the tensile string, $V^a$ was interpreted as the null eigenvector the worldsheet metric acquires in the tensionless $(T\to 0)$ limit. As $V^a$ and $v^a$ are always proportional the same interpretation holds for $v^a$.

It is not entirely clear whether the gauge-fixed BRST-invariant action \eqref{def_gauged_BRST_action} constructed for the gauged chiral string \eqref{def_gauged_bosonic_action} of this paper is automatically equivalent to what one would obtain from the first-order form of the Schild action \eqref{def_tensionless_Mink} directly, as we have not shown whether all of the necessary redefinitions can be realised as canonical transformations in the BV formalism. However, work by Sundborg \cite{Sundborg:1994py} suggests that the theory of the Schild string should be thought of as ``topological'', and its realisation as a cohomological field theory through the BRST operator defined in this paper would accord quite naturally with this suggestion.

\subsection{Tensionless Anti-de Sitter twistor strings}
A twistor action for strings (as well as particles or $p$-branes, but we will focus on strings) moving through $D$-dimensional Anti-de Sitter spacetime (AdS$_D$) was recently put forward in \cite{Arvanitakis:2016vnp}. The twistor reformulation applies only to tensionless strings, this time in the sense
\be
T R^2 \to 0
\ee
where $T$ is the string tension and $R$ is the AdS$_D$ radius. Therefore, in contrast to the Schild string, we are now setting to zero the \emph{dimensionless} quantity $(T R^2)$, as opposed to $T$. In the context of AdS/CFT, this limit corresponds to vanishing 't Hooft coupling on the CFT side (see e.g. the discussion by Tseytlin \cite{Tseytlin:2002gz}). This AdS$_D$ twistor string action is
\be
\label{def_hamiltonian_adsdivision_chiral_string_action}
S[\mathbb{Z};\rho,\mathbb{A}]=\int dt d \sigma \: \left\{\tr_\mathbb{R}\left[ \frac{1}{2}\mathbb{Z}^\dagger \Omega \partial_t\mathbb{Z} -\mathbb{A} \left(\frac{1}{2}\mathbb{Z}^\dagger \Omega \mathbb{Z}\right)\right] -\rho\,\tr_\mathbb{R}\left[ \frac{1}{2}\mathbb{Z}^\dagger \Omega \partial_\sigma\mathbb{Z} \right] \right\}\,.
\ee

The action depends on the twistor variable $\mathbb{Z}$, which is a $4\times 2$ matrix valued in a division algebra $\mathbb{K}=\mathbb{R,C,H}$ (where $\mathbb{H}$ represents the quaternions). The choice of $\mathbb{K}$ determines the dimensionality of the AdS$_D$ spacetime through $D=\dim\mathbb{K} +3$. The model has an $O(2;\mathbb{K})= O(2),U(2),{\rm Spin}(5)$ gauge invariance for $\mathbb{K}=\mathbb{R,C,H}$ respectively, which is enforced in the action by a lagrange multiplier $\mathbb{A}$ which is a $2\times 2$ $\mathbb{K}$-antihermitian matrix, on top of a more standard worldvolume spatial diffeomorphism invariance enforced by the real lagrange multiplier $\rho$. The matrix $\Omega$ is the standard $4\times 4$ symplectic matrix, and the dagger denotes $\mathbb{K}$-hermitian conjugation. We refer to \cite{Arvanitakis:2016vnp,Arvanitakis:2016wdn} for more details on this division-algebra notation.

The AdS$_D$ twistor string action \eqref{def_hamiltonian_adsdivision_chiral_string_action} is a special case of the gauged chiral string action \eqref{def_gauged_bosonic_action} where $v^a$ has been fixed to
\be
v^a=\begin{pmatrix} v^t\\v^\sigma \end{pmatrix}= \begin{pmatrix} 1\\-\rho \end{pmatrix}
\ee
and $z^A$ is identified with the real components of $\mathbb{Z}$. Since the bilinear form $\tr_\mathbb{R}[\mathbb{Z}_1^\dagger \Omega \mathbb{Z}_2]$ is antisymmetric and nondegenerate in $\mathbb{Z}_1$ and $\mathbb{Z}_2$, and since the $O(2;\mathbb{K})$ constraints have already been verified to close among themselves in \cite{Arvanitakis:2016vnp}, the considerations of the previous section imply that the gauge-fixed BRST action for the AdS$_D$ twistor string can be chosen to take the form \eqref{def_gauged_BRST_action}.

The field $\mathbb{Z}$ of the AdS$_D$ twistor string action \eqref{def_hamiltonian_adsdivision_chiral_string_action} naturally transforms as a worldsheet scalar density of weight $+1/2$ under diffeomorphisms, as does $z^A$ of the gauged chiral string for the choice $x=0$. This follows from a straightforward calculation of the gauge transformations generated from the Poisson brackets of the constraint $\tr_\mathbb{R}[\mathbb{Z}^\dagger \Omega \partial_\sigma\mathbb{Z}]$ enforced by $\rho$. In this sense, $x=0$ is singled out for this model, and no canonical transformation (in the BV sense) is necessary to make the gauge transformations of the gauged chiral string match those of the AdS$_D$ twistor string. There is thus no reason to suspect the two actions describe different theories. We note that this reformulation of the AdS$_D$ twistor string clarifies the worldsheet diffeomorphism invariance of the model, which is somewhat opaque in the original action \eqref{def_hamiltonian_adsdivision_chiral_string_action}.

\subsection{Ambitwistor strings}
\label{section_ambitwistor}
The name ``ambitwistor string'' refers to a class of models introduced by Mason and Skinner in \cite{Mason:2013sva}. The simplest version of the original model is
\be
\label{def_ambitwistor_MS}
S[X^\mu,P_\mu;e]=\int_\Sigma\left\{ P_\mu \bar\partial X^\mu - e P^2\right\}\,.
\ee
This looks similar to the Lindstrom-Sundborg-Theodoridis action for the tensionless Minkowski string \eqref{def_tensionless_Mink} which we treated above, as has already been discussed in the literature \cite{Casali:2016atr}. However there are some important differences which are relevant if we are to understand the precise relation to the gauged chiral string action \eqref{def_gauged_bosonic_action}.

The worldsheet $\Sigma$ of the ambitwistor string is assumed to be a Riemann surface with holomorphic coordinate $\sigma\in\mathbb{C}$ (with conjugate $\bar \sigma$). $\bar \partial$ then denotes $\partial_{\bar \sigma}$. The dynamical fields ($X^\mu,P_\mu,e$) appearing in the ambitwistor string action \eqref{def_ambitwistor_MS} are all complex-valued, and thus the action itself is complex, in contrast to the Lindstrom-Sundborg-Theodoridis action \eqref{def_tensionless_Mink}, the gauged chiral string action \eqref{def_gauged_bosonic_action}, and all other action functionals which have appeared in this paper so far\footnote{The AdS$_D$ twistor string action is also real. This can be seen immediately from the presence of the $\tr_\mathbb{R}[\--]$ (real trace) operation in \eqref{def_hamiltonian_adsdivision_chiral_string_action}.}. The $\bar \partial$ operator is also complex but that is less of an issue since $\bar \partial$ could arise from Wick rotating a derivative along a worldsheet lightcone direction.

To relate the ambitwistor string to our gauged chiral string one must thus consider a complexified version of \eqref{def_gauged_bosonic_action}, where $z^A$, $v^a$ and $A^I$ are all complex. Fortunately the calculations of section \ref{section_BRST} are unchanged for the complexified model. We can then consider \eqref{def_gauged_bosonic_action} in holomorphic $(\sigma,\bar\sigma)$ coordinates and set
\be
z^A=\begin{pmatrix} X^\mu\\P_\mu\end{pmatrix}\,,\quad \Omega_{AB}= \begin{pmatrix} 0 & -1 \\ 1 & 0 \end{pmatrix}\,,\quad v^a=\begin{pmatrix} v^\sigma \\ v^{\bar \sigma}\end{pmatrix}=\begin{pmatrix} 0\\1\end{pmatrix}\,,\quad A^I=2e\,,\quad  M_{IAB}z^Az^B= P^2
\ee
to obtain the bosonic ambitwistor string action \eqref{def_ambitwistor_MS}. While the actions do match after gauge fixing, the variables $z^A$ and $X^\mu,P_\mu$ transform differently under worldsheet diffeomorphisms and, much like we saw for the Schild string in subsection \ref{section_schild}, it is not clear that the BRST-invariant action \eqref{def_gauged_BRST_action} derived for the (complexified) gauged chiral string is also appropriate for this ambitwistor string.

There is another class of ambitwistor string models however, whose interpretation as gauged chiral strings does not suffer from this ambiguity. These are the ``four-dimensional ambitwistor strings'' of \cite{Geyer:2014fka}. The bosonic action of that model reads
\be
S[Z^{A'},W_{A'};a]=\int_\Sigma \left\{ W_{A'} \bar\partial Z^{A'} - Z^{A'}\bar \partial W_{A'} + a (Z^{A'} W_{A'})\right\}
\ee
where $Z^{A'} \in \mathbb{T}\cong \mathbb{C}^4$ and $W_{A'} \in \mathbb{T}^\ast$ are worldsheet spinors, i.e. transform with density weight $+1/2$, as do the $z^A$ of the gauged chiral string (with the choice $x=0$). One can thus match the two models up after identifying
\be
z^A=\begin{pmatrix} Z^{A'}\\W_{A'}\end{pmatrix}\,,\quad \Omega_{AB}= \begin{pmatrix} 0 & -1 \\ 1 & 0 \end{pmatrix}\,, \quad v^a=\begin{pmatrix} v^\sigma \\ v^{\bar \sigma}\end{pmatrix}=\begin{pmatrix} 0\\1\end{pmatrix}\,,\quad A^I=-2a\,,\quad  M_{IAB}z^Az^B= Z^{A'} W_{A'}\,.
\ee

An interesting point arises if we consider the Weyl transformations of the complexified chiral string model (before the gauge fixing we described just now), which send $(Z^{A'},W_{A'})\to (\omega Z^{A'}, \omega W_{A'})$ where the local scaling $\omega=\omega(\sigma)$ is now complex-valued. At the same time, the constraint $Z^{A'} W_{A'}=0$ will also generate a local scaling, where now $Z$ and $W$ transform oppositely, i.e. $(Z^{A'},W_{A'})\to (\omega' Z^{A'}, (\omega')^{-1} W_{A'})$. After we quotient by both transformations, and ignoring the fact that the $\omega$ transformation also acts on $v^a$, we find that the target space of the model is the locus $Z^{A'} W_{A'}=0$ in $\mathbb{PT}\times \mathbb{P T^\ast}$, i.e. projective ambitwistor space.

\subsection{Others}
By now it should be clear how string actions with ultralocal, quadratic constraints can be seen as special cases of the (gauged) chiral string action \eqref{def_gauged_bosonic_action}. For two further examples, we point out the $SU(2,2)$-invariant two-dimensional model of Claus, Gunaydin, Kallosh, Rahmfeld and Zunger \cite{Claus:1999xr} (which can be obtained from the ungauged chiral string, or equivalently from the AdS$_5$ tensionless twistor string (i.e. the one with $\mathbb{K}=\mathbb{C}$) described above by removing the $U(2)$ constraints), and the Hohm-Siegel-Zwiebach string \cite{Hohm:2013jaa}, at least in its ``halved'' form (formula (2.16) of that paper).

\section{Discussion}
\label{section_Discussion}
We have introduced a sigma model action \eqref{def_S0} describing maps from a $p$-brane worldvolume into a symplectic target space. Besides its relevance for a number of previously considered theories, this action is interesting for some of its technical features: its Weyl invariance acting on both target space as well as the worldvolume vector field $v^a$, its diffeomorphism invariance in spite of the absence of a worldvolume metric and the closely related fact that it can apparently be written as a total BRST variation (after the ghost sector is introduced).

The last claim is subject to technical caveats detailed in subsection \ref{subsection_unbelievers}. To summarise that discussion: it is true that there is a choice of off-shell nilpotent BRST operator (``BRST2'', defined in \eqref{def_BRST_modified} for the ungauged model and explicitly by formulas \eqref{BRST_z}, \eqref{BRST_ca}, \eqref{BRST_comega}, \eqref{BRST_ba}, \eqref{BRST_cI} and \eqref{BRST_bI} in the general case) such that the action is BRST-trivial, but we have not been able to verify that this remains true after some ghost variables are eliminated. The choice of ghost fields to be integrated over in the path integral is closely related to the choice of measure and as such, this issue would be clarified when the quantum theory of the chiral brane model is formulated, perhaps along the lines of \cite{Dedushenko:2010zn}.

Assuming that the BRST operator we define is in fact appropriate, the chiral brane theory is an example of a cohomological field theory (also known as ``topological''). Such theories enjoy a number of special properties. To see some of them, let us briefly review an argument due to Witten \cite{Witten:1988xj} (reviewed in \cite{Witten:1990bs} and \cite{Vonk:2005yv}) that shows the semi-classical approximation to the path integral for such theories is exact. Let us assume the absence of BRST anomalies. In the path integral formulation this is equivalent to assuming the measure is BRST invariant \cite{Fujikawa:1979ay,Fujikawa:1980eg}. If we also assume that the gauge-fixed action $S$ entering the path integral is BRST invariant, invariance of the measure can be equivalently expressed as
\be
\forall \mathcal{O}:\qquad\langle \delta_{\rm BRST}\mathcal O\rangle \equiv\int \mathcal{D}\phi\: (\delta_{\rm BRST}\mathcal O)  \exp(i S/\hbar)=0\,.
\ee
Therefore, if $\mathcal V$ denotes any product of \emph{BRST-invariant} operators (i.e. $\delta_{\rm BRST} \mathcal V=0$), while $\mathcal O$ is still an arbitrary operator, we must have
\be
\forall \mathcal{O}:\qquad\langle \mathcal V \,\delta_{\rm BRST} \mathcal O \rangle = 0\,.
\ee
Now consider varying any correlation function $\langle \mathcal V \rangle$ with respect to $\hbar$. If $S=\delta_{\rm BRST} B$ as is the case for the chiral brane then
\be
\frac{\partial}{\partial \hbar} \langle \mathcal V \rangle=-i\hbar^{-2}\langle\mathcal V \delta_{\rm BRST} B\rangle =0\,.
\ee
Therefore, assuming these formal manipulations are not obstructed (by e.g. the nonexistence of an appropriate path integral measure), the theory is independent of the value of $\hbar$ and we can calculate in the limit $\hbar \to 0$, where the path integral localises on the solutions of the equations of motion. A similar argument implies the theory cannot (continuously) depend on the gauge choice $\tilde v^a$. Since a worldvolume metric only enters the theory through that gauge choice it should follow that the theory also does not depend on that. Some of these properties were anticipated in \cite{Casali:2016atr} (for the ambitwistor string).

Given that the Schild string \eqref{def_tensionless_Mink} and the AdS$_D$ twistor string \eqref{def_hamiltonian_adsdivision_chiral_string_action} actions were both derived from tensionless limits of the usual tensile string, the above properties of the BRST operator suggested in this paper appear sensible: after all the string tension, or more accurately its inverse $\alpha'\propto T^{-1}$, plays the role of $\hbar$, but as we have already taken the limit $\alpha' \to \infty$ in deriving those actions, it would be strange if another loop-counting parameter somehow appeared in these theories. It is thus fortunate that these theories turn out to be independent of the value of the prefactor in the action. Note that we are referring to worldsheet rather than target space loops here: it should be possible to obtain target space loop corrections by considering worldsheets of arbitrary genus. In effect what the previous argument shows is that there is no $\alpha'$ expansion, but there might still be a $g_S$ 
expansion. Of course these conclusions will only follow if 
the quantum theories are critical (i.e. no BRST anomalies), which will only be true in certain dimensions and/or for some particular number of supersymmetries. For this reason it would be interesting to consider the supersymmetric generalisation of the chiral brane action.

On the other hand, it is not clear whether the choice of BRST operator proposed in this paper is appropriate for ambitwistor strings for a number of reasons. For one, there appears to be no consensus in the literature on whether the diffeomorphisms of the ambitwistor model are gauged \cite{Lipstein:2015vxa}, whereas the diffeomorphisms of the chiral string introduced in this paper are, of course, gauged. For another, the argument above is only valid in the absence of BRST anomalies, i.e. in the critical dimension, while ambitwistor strings are often considered outside of their critical dimension (e.g. in \cite{Geyer:2014fka}). Ambitwistor string amplitudes do seem to enjoy certain localisation properties however and it would be natural to assume this localisation is realised through the mechanism described in this paper.

\section*{Acknowledgements}
This work was prompted by an interesting discussion involving Alec Barns--Graham, David Skinner, and Jack Williams.

I would like to thank Kai Roehrig for a clarification regarding ambitwistor strings, Kostandinos Sfetsos and the Faculty of Physics at the University of Athens for their kind hospitality, and Paul Townsend for discussions. I would also like to thank Kenny Wong for bringing the review \cite{Vonk:2005yv} to my attention.

\appendix

\section{BRST transformations}
The BRST transformations under which the gauge-fixed action \eqref{def_gauged_BRST_action} for the gauged chiral string is invariant are
\begin{align}
\label{BRST_z}
\delta_{\rm BRST} z^A &=\left(c^b_\xi \partial_b z^A + \frac{1}{2} \partial_b c^b_\xi z^A - c_\omega z^A + c^I M_{IBC}\Omega^{AB} z^C \right) \Lambda\\
\label{BRST_ca}
\delta_{\rm BRST} c^a &= (- c^b_\xi \partial_b c^a_\xi) \Lambda\\
\label{BRST_comega}
\delta_{\rm BRST} c_\omega&= (- c^b_\xi \partial_b c_\omega)\Lambda \\
\label{BRST_ba}
\delta_{\rm BRST} b_a &=\left( - \frac{1}{2} \Omega_{AB} z^A \partial_a z^B + \partial_b(b_a c^b_\xi) + b_b \partial_a c^b_\xi - 2 b_a c_\omega + b_I \partial_a c^I\right)\Lambda\\
\label{BRST_cI}
\delta_{\rm BRST} c^I &= \left( - \frac{1}{2} f_{JK}{}{}^I c^J c^K - c_\xi^a \partial_a c^I \right)\Lambda\\
\label{BRST_bI}
\delta_{\rm BRST} b_I &= \left( \frac{1}{2} M_{IAB} z^A z^B - f_{IJ}{}{}^K c^J b_K - c^a_\xi \partial_a b_I - \partial_a c^a_\xi b_I + 2 c_\omega b_I \right)\Lambda\\
\delta_{\rm BRST} \tilde v^a &=0\,.
\end{align}
$\Lambda$ is the constant, anticommuting BRST transformation parameter of ghost number $\gh \Lambda=-1$.

In the ungauged case, all terms with $IJK$ indices are absent. The BRST transformations for the gauged model were constructed by perturbing those of the ungauged one by quadratic terms involving $IJK$ indices, and then demanding that the resulting BRST transformation squares to zero \emph{off-shell}, i.e.
\be
\delta_{\rm BRST}^2\equiv\delta_{{\rm BRST};\Lambda_2} \delta_{{\rm BRST};\Lambda_2}\phi =0
\ee
for any $\Lambda_1,\Lambda_2$ and where $\phi$ is any of the above fields. The vanishing of the left-hand side fixes the coefficients in the BRST variation. The values for those coefficients were obtained with the help of the computer algebra programme Cadabra (v. 1.39) \cite{DBLP:journals/corr/abs-cs-0608005,Peeters:2007wn}, and the off-shell nilpotence of the BRST variation derived thereby was subsequently verified by hand.

More precisely: we considered the ansatz
\begin{align}
\delta_{\rm BRST} b_a &=\left( - \frac{1}{2} \Omega_{AB} z^A \partial_a z^B + \partial_b(b_a c^b_\xi) + \partial_b \partial_a c^b_\xi - 2 b_a c_\omega +\bm{\theta} b_I \partial_a c^I\right)\Lambda\\
\delta_{\rm BRST} c^I &= \left(  \frac{1}{2}\bm{\kappa} f_{JK}{}{}^I c^J c^K +\bm{\alpha} c_\xi^a \partial_a c^I + \bm{\beta} c_\omega c^I + \bm{\gamma} \partial_a c^a_\xi c^I \right)\Lambda\\
\delta_{\rm BRST} b_I &= \left( \frac{1}{2}\bm{\mu} M_{IAB} z^A z^B +\bm{\lambda}f_{IJ}{}{}^K c^J b_K +\bm{\zeta} c^a_\xi \partial_a b_I +\bm{\epsilon} \partial_a c^a_\xi b_I + \bm{\eta} c_\omega b_I \right)
\end{align}
where the variations for the other fields are as above and the real parameters $\{\bm{\alpha,\beta,\gamma,\epsilon,\zeta,\eta,\theta,\kappa,\lambda,\mu}\}$ were determined along with the relations between the constant tensors $M_{IAB},\Omega_{AB},\Omega^{AB}$ and $f_{IJ}{}{}^K$ (assumed to have the (anti)symmetry properties described in the text). We then found
\begin{align}
\delta_{\rm BRST}^2 z^A =0 &\iff \bm{\alpha}=-1\,, \bm{\beta}=0\,, \bm{\gamma}=0\,, \left(\frac{1}{2}\bm{\kappa} M_{KAB} f_{IJ}^{}{}^K+ M_{I A C}\Omega^{CD} M_{J BD}\right) \Omega^{EA}=0\,,\\
\delta_{\rm BRST}^2 c^I=0 &\implies f_{[IJ}{}{}^K f_{L] K}{}{}^M=0\,,\\
\delta_{\rm BRST}^2 b_I=0 &\implies \bm{\epsilon}=-1\,,\bm{\zeta}=-1\,, \bm{\eta}=-2\,, \bm{\lambda}=\bm{\kappa}\,,\\
\delta_{\rm BRST}^2 b_a =0 &\implies \Omega^{AC}\Omega_{BC}=+\delta^A_B\,,\bm{\mu\theta}=1\,,
\end{align}
where in deriving each implication we have already substituted in the values of the parameters fixed above. The parameters $\bm{\kappa}$ and either $\bm{\mu}$ or $\bm{\theta}$ are arbitrary, and were fixed to $\bm{\kappa}=-1,\bm{\theta}=\bm{\mu}=1$ to produce the transformations above. The upshot is that the BRST transformations are nilpotent if conditions \eqref{BRST_conditions} on $M_{IAB},\Omega_{AB},\Omega^{AB}$ and $f_{IJ}{}{}^K$ are satisfied.

\bibliography{Bib}
\bibliographystyle{JHEP}

\end{document}